\documentstyle[aps,prd,12pt]{revtex}
\begin{document}
\pagestyle{empty}
\begin{center}
{{\Large \bf Cosmological Constant in de-Sitter Spacetime} \footnote {to appear in the Proceedings, IMFP'98, Oct.26-29, Malaysia, Kualalumpur}}
\end{center}
\vspace {0mm}
\begin{center}
 Ishwaree P. Neupane \footnote{Electronic mail: ishwaree@phya.snu.ac.kr}\\
{Department of Physics,Seoul National University, Seoul 151-742, Korea}
\footnote{ On leave from {\it Department of Physics, Tribhuvan University, Kirtipur, Kathmandu, Nepal}} 
\end{center}
\vspace {0mm}
\begin{abstract}
With the basic cosmological relations that agree with the recent observations,
simple expressions are suggested concerning the value of cosmological 
constant($\Lambda$). A large contribution of quantum vacuum to the energy 
momentum tensor does not agree with the observed cosmos. However,one requires
the presence of positive $\Lambda$ to make the various observations consistent. After a review of the effect of cosmological constant on the geodetic motions in the Schwarzschild de Sitter spacetime, some approaches to its solutions are briefly discussed. Also suggested is the very weak limit on $\Lambda$ from the  planetary perturbations. 
\end{abstract}
\section{Introduction}
The cosmological constant($\Lambda$) has been an outstanding problem for the 
past seventy-five years[1,2], ever since Einstein introduced it in the field
equations to avoid an expanding universe. One of the great developments of 
1980's, was the creation of a 
standard model of cosmology based on the ideas arising from particle physics. This model involved the following trilogy of ideas: (i) $\Omega$=1, 
(ii) $\Lambda=0$ and (iii) $\Omega_{matter}\approx\Omega_{{CDM}^{WIMP}_{axion}}
\geq0.9$. But, such a model of 1980's is no more[1]. Infact, the density of the
matter insufficient to result in a flat universe($\Omega=1$) suggests a positive $\Lambda$. One would now prefer either (1) $\Omega\neq1$, $\Lambda=0$, 
$\Omega_{matter}\approx\Omega_{{CDM}^{WIMP}_{axion}}\approx0.1-0.3$ or
(2) $\Omega\equiv1$, $\Lambda\neq0$, 
$\Omega_{matter}\approx\Omega_{{CDM}^{WIMP}_{axion}}\approx0.1-0.3$. 
A small non-vanishing $\Lambda$ is required to make the two independent
observations: the Hubble constant ($H_o$ : which explains the expansion rate of
the present universe) and the present age of the universe $(t_o)$ consistent each other[3]. This has forced us critically re-examine the simplest and most 
appealing cosmological model- a flat universe with $\Lambda=0$[3,4]. A flat 
universe with $\Omega_{m}\equiv0.3$ and $\Omega_{m} + \Omega_{\Lambda}=1$ is
most preferable and a matter dominated flat universe with $\Lambda=0$ is ruled 
out[5].\\
A various inflationary models suggest that the scale factor $a(t)$ of the universe after inflation has been in the order of $\sim10^{28}$ before the inflation, which created a smooth and effectively flat universe of the right size and entropy. This further implies that the inhomogeneity is created by quantum fluctuation of the scalar field in the inflationary phase and gives us a preference for $\Omega=1$ and claims $\Lambda\neq0$[6]. In fact $\Lambda$ follows from the dynamical evolution of our universe as one interprets it as the vacuum energy of the quantized fields[7]. Today $\Lambda$ has incredibly small value $\sim10^{-47} (GeV)^4$[3,6], while the quantum field theories in curved spacetime would imply quite different values of vacuum energy density($\rho_v$) in the early universe (in units ${8\pi G=1}$ ,we often denote ${\rho_v}$ by ${\Lambda}$). The generic inflation models also require it to have a large value during the inflationary epoch[6]. In particular, $\Lambda_{GUT}\sim10^{64}(Gev)^4$, $\Lambda_{EW} \sim 10^8 (GeV)^4$ and $\Lambda_{QCD} \sim10^{-4} (GeV)^4$[8]. This gives, up to some extent, a natural explanation for the small value of $\Lambda$ at present and its large value in the early universe. Also, a high intensity of vacuum energy in the inflation era (i.e.,de Sitter phase) corresponds to the large
vacuum energy needed to drive inflation[9]. However, it is unlikely that such quantum instabilities can lower the value of $\Lambda$ by a large factor and yield a universe even remotely like our own[10]. The problem of cosmological constant is therefore to explain these huge orders of magnitude difference between $\rho^{early}_{vac}$ and $\rho^{p}_{vac}$ in a natural manner, in particular without fine tuning of the values of cosmological parameters[8]. A number of time dependent cosmological constant models, such as $\Lambda\propto (1/t^2)$; t as the age of present universe suggest that the vacuum energy is a function of scalar(dilatonic) field, an idea supported also from non criticle string theories[2].\\

In this note we consider the above facts on the basis of simple cosmological 
relations and recent determinations of cosmological parameters. We explain a 
case of non zero $\Lambda$ by considering its effect on the geodetic motions in the Schwarzschild de-Sitter spacetime.\\
\section{The Standard Cosmology and $\Lambda$}
To a good approximation the present universe is spatially homogeneous and isotropic on large scales[11]. So the space-time geometry of the universe is appropriately described by Robertson-Walker(RW) metric in the form
\begin{eqnarray}
ds^2=dt^2-a^2(t) \Big[ \frac{dr^2}{1-kr^2}+r^2 d\theta^2 + r^2\sin^2\theta d\phi^2 \Big]
\end{eqnarray}
where $a(t)$ is the cosmic scale factor and $k=+1, 0, -1$ depending upon whether the universe is closed, flat or open. The vacuum expectation value of energy momentum tensor of quantum fields in de Sitter space takes the form[12]
$<T_{\mu\nu}^{vac}>=\rho_{vac} g_{\mu\nu}$. So a model universe with an additional term $\rho g_{\mu\nu}$ in the Einstein field equation is highly motivated and the $\Lambda$ corresponding to the vacuum energy density enters in the form
\begin{eqnarray}
R_{\mu\nu} - \frac{1}{2}g_{\mu\nu}R =G_{\mu\nu}= 8\pi GT_{\mu\nu} - \Lambda g_{\mu\nu}
\end{eqnarray}
which with eqn(1) gives the so-called Friedmann equation
\begin{eqnarray}
\Big( \frac{\dot{a}}{a} \Big)^{2} = \frac{8\pi G}{3}\rho - \frac{k}{a^{2}} + \frac{\Lambda}{3}
\end{eqnarray}
where $\rho$ is the energy density of baryonic plus dark matter. Eqn(3) gives 
\begin{eqnarray}
\frac{k}{{H_o}^{2}{a_o}^{2}}=\Omega_{m}+\Omega_{v} - 1
\end{eqnarray}
where $\Omega_m=\rho_m/\rho_c$ and $\Omega_v=\rho_v/\rho_c$ are respectively  the matter and vacuum density parameters; and $\rho_{c}=3{H_o}^{2}/8\pi G$ and $\rho_{v}=\Lambda/8\pi G$.\\

A number of recent observations would converge the present value of the Hubble constant in the range: (1) $H_{o} = 67\pm 6{} kms^{-1} Mpc^{-1}$ ( Nevalainen and Roos '97), (2) $H_{o} = 70\pm 5{}kms^{-1} Mpc^{-1}$ (Freedmann '98). Since $H_{o}^{2} a^{2} \sim 1$, eqn(4) gives the following relations:
(i) for $k=0,{}\Omega _{m} + \Omega _{v}=1$, (ii) for $k=1,{}\Omega _{m} + \Omega _{v} = 2$ and (iii) for $k=-1,{}\Omega _{m} + \Omega _{v} = 0$. However, the present observational limit allows $0.2<\Omega <2$. The present energy density contributed by matter is estimated as $\Omega= 0.1\sim 0.4$ in a broad range[13] which corresponds to $\rho_v \leq 10^{-47} (GeV)^4$.\\
\section{The $\Lambda_{eff}$ and Age of the  Universe}
The limit on the present age of the universe taken from the age of the oldest clusters corresponds to $t_{globulars} = 11.5\pm 1.3 Gyr$ (C.Hipparcos et.al.'97); to which the age of the universe at the time of their formation must be added; while the dynamical age of the universe would be $t_{o} = 14.2\pm 1.5 Gyr$, which includes the systematic uncertainties in the current cepheid distance scale[14]. 
According to Friedman-Lamaitre model, the age $t(z)$ of the universe at 
redshift $(z)$ is expressed by
\begin{eqnarray}
t(z) = \frac{1}{H_o}\int_{0}^{1/1+z} dx \Big[(1-\Omega_m-\Omega_{\Lambda})+\Omega_m x^{2-3(1+w)}+\Omega_{\Lambda}x^2 \Big]^{1/2}
\end{eqnarray}
with the equation of state $p=w\rho$. With the given values of $H_o$ and $t(z)$, the above equation puts contraint on $\Omega_m$ and $\Omega_{\Lambda}$. For $w=0 ( i.e., p<<\rho)$; and $H_o = 60-75 kms^{-1}Mpc^{-1}$ and $t_o = 12.8{}Gyr$, one gets $\Omega_m=0.2-0.4$ for $\Omega_{\Lambda}=0.6-0.8$. However, the statistics of gravitational lensing puts the upper limit $\Omega_{\Lambda}\leq 0.66$ for a flat universe ( Kochanek '96 ), while the observations of the clusters of galaxies put the lower limit $\Omega_{\Lambda}\geq0.6$[15]. 
Based on eqn(5) the limits $\Omega _m < 0.22$ and $\Omega _{\Lambda} > 0.6$ are indicated in the ref.[16]. Therefore, $\Omega _{\Lambda} = 0.6 - 0.66$ puts a very strong limit on $\Lambda_o$ to be $(2\sim3)10^{-47}(Gev)^4$ for 
$\Omega_{vac} + \Omega_{matter} = 1$.\\
The age of a flat universe that contains both matter and positive vacuum energy
is given by[11]
\begin{eqnarray}
t_o = \frac{2}{3}H_{o}^{-1}\Omega_{vac}^{-1/2}  \ln \Big[ \frac{1+\Omega_{vac}^{1/2}}{(1-\Omega_{vac})^{1/2}} \Big]
\end{eqnarray}
A value of $H_o$ to its lower edge (e.g. $64 kms^{-1} Mpc^{-1}$) and the value of $t_o = 14.2  Gyr$ implies $H_o t_o = 0.93$ which corresponds to 
$\Omega _{vac} = 0.66$. For $\Omega _{vac} = 0.7$, one gets $H_o t_o = 0.964$. So $t_o=14.2Gyr$ implies $H_o=66{}kms^{-1}Mpc^{-1}$. For $\Omega_{vac} = 0.8$, one gets $H_o t_o = 1.076$. A model universe with $\Omega_{vac}\geq 0.74$ is older than $H_{o}^{-1}$, thereby implies an accelerating universe and in the limit $t_o \to\infty$ one gets $\Omega_{vac}\to1$(i.e., a de-Sitter solution). Also, the higher values of $\Omega_{vac}$ start to conflict with the lower bound on matter energy density from galaxies and clusters. So we usually discard the value $\Lambda>0.74$ based on the ideas from the experimental evidences.\\
%%%%%%%%%%%%%%%%%%%%%%%%%%%%%%%%%%%%%%%%%%%%%%%%%%%%%%%%%%%%%%%%%%%%%
\section {Schwarzschild de-Sitter Spacetime and $\Lambda$}
In this section we express $\Lambda$ in the unit of $cm^{-2}$. So we define 
$\Lambda_{pl}=\Lambda (a/l_{pl})^{\alpha}, (a/l_{pl})$ is the scale factor in units of Planck length, $\Lambda_{pl}\sim M_{pl}^2$ is the natural size of the cosmological constant and $\alpha$ is a constant to be determined by the present upper bound on $\Lambda$. Since $a_o/l_{pl}\sim ct_o/l_{pl}\sim 10^{61}$, the present upper bound on $\Lambda_{o}\leq 10^{-123} M_{pl}^2$ implies $\alpha\equiv2$ and therefore a relation $\Lambda_{pl}\equiv\Lambda (a/l_{pl})^2$ is claimed to explain the spontaneous decay of $\Lambda$ from its large value at Planck's era to its extremely small value at the present universe. In this formalism the value $\Lambda_{o}\leq 10^{-47}(GeV)^4$ corresponds to $|\Lambda_o|\leq 10^{-123} M_{pl}^{2}\approx 10^{-56} cm^{-2}$.\\

A generally spherically symmetric metric is described by the form
\begin{eqnarray}
d\tau^2 = e^{2\lambda(r,t)} dt^2 - e^{2v(r,t)} dr^2 - r^2 (d\theta^2 + \sin^2 d\phi^2)
\end{eqnarray}
Corresponding to the vacuum field equations $G_{\mu\nu}=-\Lambda g_{\mu\nu}$, the generalization of the Schwarzschild solution for the above metric by allowing non-zero cosmological constant [17,18](in units $c=G=1$) is given by
\begin{eqnarray}
d\tau^2 = \Big( 1-\frac{2M}{r}-\frac{\Lambda r^2}{3} \Big)dt^2 - \frac{dr^2}{\displaystyle{1-\frac{2M}{r}-\frac{\Lambda r^2}{3}}} - r^2 (d\theta^2 + \sin^2 \theta d\phi^2)
\end{eqnarray}
The above metric is considered as the Schwarzschild de-Sitter metric and hence the space determined by it is not asymtotically flat as in the case of Schwarzschild metric, for $\Lambda$ related to the vacuum energy density implies a pre-existing curvature[18]. It is easy to see that the Lagrangian and Hamiltonian for this metric are equal. So there is no potential energy in the problem. By rescaling $\tau$ and setting $\theta=\pi/2$ ( i.e., an equatorial plane), we get
\begin{eqnarray}
\frac{E^2}{\displaystyle{1-\frac{2M}{r}-\frac{\Lambda r^2}{3}}} - \frac{\dot{r}^2}{\displaystyle{1-\frac{2M}{r}-\frac{\Lambda r^2}{3}}} - \frac{L^2}{r^2} = 2{\cal L}= +1 ~\mbox{or} ~0
\end{eqnarray}
for the time like or null geodesics respectively; where $E=(1-2M/r-\Lambda r^2/3)\dot{t}$ and $L=r^2 \dot{\phi}$ are the constants associated with the energy and momentum of the particles respectively.
In the case of physical interest (i.e.,for the time like geodesics), considering $r$ as a function of $\phi$ and letting $u=1/r$, we get
\begin{eqnarray}
\frac{d^2 u}{d\phi^2} + u = \frac{M}{L^2} + 3Mu^2 - \frac{\Lambda}{3L^2 u^3}
\end{eqnarray}
The numerical solutions to this quintic polynomial with some constraints, e.g.  $E^2+\Lambda L^2/3<1$ and $\Lambda M^2<1/9$ for bound orbits, can show only three real roots with two positive and one negative. Out of these
roots, the two smaller roots would be near the cosmological horizon and the larger root would be near the black hole horizon, and no real roots are present in the region $r_{CO}<r<r_{BH}$[19]. There could be two more roots, but they are essentially imaginary. As the function is negative in the region $r_{CO}<r<r_{HB}$, finding the exact analytic solutions in terms of the elliptic integrals with all roots seems much more complicated than the case in Schwarzschild metric. However, to study the effect of $\Lambda$ in the geodetic motion of the particles one can treat the third term on rhs of eqn(10) as a perturbation, for it is $\sim 10^{-8}$ of the first term and $\sim 10^{-6}$ of the second term in the case of Mercury's orbit with $\Lambda \approx 10^{-43} cm^{-2}$.
A simple approximation to the problem shows that the main effect of the term involving $\Lambda $ in eqn(9) is to cause an additional advance of the perihelian by an amount (retaining all the parameters in their original units)
\begin{eqnarray}
\Delta\phi_{\Lambda} = \frac{\pi\Lambda c^2 a^3 (1-e^2)^3}{GM}=\frac{2\pi\Lambda l^3}{r_s}
\end{eqnarray}
where $a$ is semimajor axis, $e$ is the eccentricity, $l$ is the semilatus rectum and $r_{s}=2GM/c^2$ is Schwarzschild radius. The additional precession due to $\Lambda$ is therefore $\Delta\phi_{\Lambda}=\pi \Lambda c^2 r^3/GM $.\\
For circular orbits, $\Delta\phi_\Lambda=\pi\Lambda c^2 a^3/GM$. If we define $\bar{\rho}$ as the average density within a sphere of radius $a$ and $\rho_{vac}=\Lambda c^2/8\pi G$ as the vacuum density equivalent of the cosmological constant, we get $\Delta\phi_{\Lambda} =  6\pi\rho_{vac}/\bar{\rho} $ rad/orbit.
For the case of bound orbits, a relation between the cosmological constant and the minimum orbit radius can be expressed by $r_{min}=(3MG/\Lambda c^2)^{1/3}$.\\
\section{Discussion}
Though, the microscopic theories of particle physics and gravity suggest a large contribution of vacuum energy to energy momentum tensor, all observations to date show that $\Lambda$ is very small and positive. In the case of Mercury, the extra precession factor $\Delta$ would be $0.1''$ per century (i.e.,the maximum uncertainty in the precession of the perihelion), if $\Lambda\leq 3.2\times 10^{-43} cm^{-2}$. With the current value of 
$|\Lambda|\leq 10^{-56} cm^{-2}$, for Mercury one gets  $\Delta\phi_{\Lambda}\leq 3.6\times 10^{-23}$ arc second per orbit; which is unmeasurably small and very far from the present detectable limit($3\times 10^{-4}$arc second) of VLBI(Very Long Baseline Interferometry). It sounds more logical to argue that only the tests based on large scale geometry of the universe can put a strong limit on the present value of $\Lambda$.\\

However, the precession in the perihelion of the planets provide a sensitive solar test for a cosmological constant. Also, for very massive binary star systems such as Great Attractor(GA) and Virgo Cluster with highly eccentric orbits, the value of cosmological constant may show up. In this case, however, an accurate profile of infall velocities of galaxies into the GA is needed to provide a good estimate of present bound on cosmological constant. For example, in the case of Pluto with $\Lambda\leq 10^{-56} cm^{-2}$, one gets $\Delta\phi_{\Lambda}=3.5\times 10^{-17}$ arc second per orbit; which is also unmeasurably small. An extremely small value of the  $\Lambda$ makes us unable to measure the extra precession with the required precision. It is here worthnoting that $\Lambda$ must be quite larger than $10^{-50}cm^{-2}$ to observe its effects possibly with an advance of additional precession of perihelion orbit in the inner planets. In the case of Pluto with $\Delta\phi_{\Lambda}\leq 0.1$ arc second per orbit, we get $\Lambda\leq 3.3\times10^{-49}cm^{-2}$, which may be very near to the bound on the present value of cosmological constant, i.e., $0\leq|\Lambda_o|\leq 2.2\times 10^{-56}cm^{-2}$. However, the planetary perturbations cannot be used to limit the cosmological constant.\\

{\large\bf Acknowledgements}\\

I would like to thank the organizers of the IMFP'98 for kind hospitality and providing partial travel support to attend the meeting.\\

{\large\bf References}
\begin{description}
\item{[1]} L.M.Krauss, {\it Preprint, hep-ph/9810393}(1998).
\item{[2]} J.Z.Lopez and D.V.Nanopoulos, {\it Mod.Phys.Lett.} A11 (1996) 1. 
\item{[3]} Anup Singh, {\it Phys.Rev.} D52 (1995) 6700.
\item{[4]} W.L.Freedmann et.al., {\it Nature}, 371(1994)757.
\item{[5]} M.Chiba and Y.Yoshii, {\it Preprint, astro-ph/9808321}, (1998).
\item{[6]} A.Linde, {\it Particle Physics and Inflationary Cosmology}, harwood academic publishers (1990).
\item{[7]} N.D.Birrel and P.C.W.Davies, {\it Quantum Fields in Curved Scape}, Cambridge Univ. Press (1982).
\item{[8]} S.Weinberg, {\it Rev.Mod.Phy.} 61(1989)1.
\item{[9]} J.W.Moffat, {\it Preprint, astro-ph/9606071} (1996).
\item{[10]} W.A.Hiscock, {\it Phys.Lett.}166B, \underline{Vol.3} (1986)285.
\item{[11]} E.W.Kolb and M.S.Turner, {\it The Early Universe} (1990).
\item{[12]} L.Ford, {\it Phys.Rev.} D28(1983)710.
\item{[13]} M.Ozer and M.O.Taha, {\it Mod.Phys.Lett.} A13 (1998) 571-580.
\item{[14]} Reiss et.al., {\it Preprint, astro-ph/9805201} (1998).
\item{[15]} Mellier et.al., {\it Preprint, astro-ph/9609197} (1996).
\item{[16]} M.Roos and S.M.Rashid, {\it Astron.Astrophys.} 329 (1998) L17.
\item{[17]} C.Kahn and F.Kahn, {\it Nature} 257 (1975) 451.
\item{[18]} G.Gibbons and S.Hawking, {\it Phys.Rev.} D15 (1977) 2738.
\item{[19]} J.Pokharel and U.Khanal, {\it Geodesics in Schwarzschild de-Sitter Spacetime}, an unpublished work, Dept. of Phys.,Tribhuvan Univ.(1997).\\
\end{description}
\end{document}